\DeclareUnicodeCharacter{2212}{\textendash}
\documentclass[12pt]{iopart}

\usepackage[numbers,comma,square,sort&compress,merge]{natbib} 
\usepackage{color,soul}
\usepackage{todonotes}
\usepackage{graphicx}

\usepackage{textcomp}
\usepackage{xspace}
\usepackage[innercaption,ragged]{sidecap}
\usepackage{url}
\usepackage{eurosym}
\usepackage{color,soul} 

\usepackage{hyperref}
\hypersetup{pdftitle={Properties of group-III-assisted (In,Ga)As nanowires grown by MBE}, colorlinks, citecolor=blue}

\begin{document}
	\title[Properties of group-III-assisted (In,Ga)As nanowires grown by MBE]{Composition and optical properties of (In,Ga)As nanowires grown by group-III-assisted molecular beam epitaxy}

	\author{M Gómez Ruiz$^{*}$, A Castro, J Herranz, A da Silva, A~Trampert, O Brandt, L Geelhaar, and J Lähnemann}
	\address{Paul-Drude-Institut für Festkörperelektronik, Leibniz-Institut im Forschungsverbund Berlin e.V., 
		Hausvogteiplatz 5–7, 10117 Berlin, Germany}
	
	\ead{$^{*}$gomez@pdi-berlin.de}
	\vspace{10pt}

	\begin{abstract}
		(In,Ga) alloy droplets are used to catalyse the growth of (In,Ga)As nanowires by molecular beam epitaxy on Si(111) substrates. The composition, morphology and optical properties of these nanowires can be tuned by the employed elemental fluxes. To incorporate more than 10\% of In, a high In/(In+Ga) flux ratio above 0.7 is required. We report a maximum In content of almost 30\% in bulk (In,Ga)As nanowires for an In/(In+Ga) flux ratio of 0.8. However, with increasing In/(In+Ga) flux ratio, the nanowire length and diameter are notably reduced. Using photoluminescence and cathodoluminescence spectroscopy on nanowires covered by a passivating (In,Al)As shell, two luminescence bands are observed. A significant segment of the nanowires shows homogeneous emission, with a wavelength corresponding to the In content in this segment, while the consumption of the catalyst droplet leads to a spectrally-shifted emission band at the top of the nanowires. The (In,Ga)As nanowires studied in this work provide a new approach for the integration of infrared emitters on Si platforms.
	\end{abstract}
	
	%
	\submitto{\NT}
	%
	\maketitle
	%
	\ioptwocol

	\section{Introduction}
	The ternary alloy In$_{x}$Ga$_{1-x}$As has a widely tunable, direct band gap ($E_{\mathrm{g}}=0.35$--1.42~eV), and thus the range of emission energies overlaps with the second and third low-loss telecommunication bands. Hence, this material is attractive for the photonic integration of emitting heterostructures on Si waveguides. To achieve the integration on Si, high quality epilayers with a low density of defects are required \cite{papanicolaouLatticeMismatchedInGaAs1993}. However, at an In content $x = 0.5$, In$_{x}$Ga$_{1-x}$As has a lattice mismatch of 8\% with silicon \cite{papanicolaouLatticeMismatchedInGaAs1993}. Together with the difference in thermal expansion coefficients \cite{bisaroThermalExpansionParameters1979}, the inevitable plastic strain relaxation leads to the emergence of misfit dislocations (MDs) in the proximity of the interface during epitaxial growth \cite{kayaStudiesLatticeMismatch2004}, giving rise to threading dislocations (TDs) extending from the interface to the epilayer surface \cite{mathisThreadingDislocationReduction1999}. The three-dimensional nanowire (NW) geometry acts as an efficient dislocation filter as the few TDs that form at the interface are bend and terminated at the lateral nanowire surfaces leaving the bulk of the NW free from dislocations\cite{Hersee_2011, borg_2014}.

	The growth of (In,Ga)As NWs has been studied both by metal organic vapour phase epitaxy (MOVPE) and molecular beam epitaxy (MBE) \cite{koblmullerGrowthPropertiesInGaAs2014}. MOVPE growth, which achieves uniform NW arrays, usually relies on an extrinsic metal droplet such as Au serving as catalyst. The contamination by this element induces non-radiative recombination centres \cite{breuerSuitabilityAuSelfAssisted2011}. Furthermore, Au easily diffuses on and into silicon and is therefore incompatible with Si technology \cite{jacksonIntegratedSiliconNanowire2007}. Using MBE, (In,Ga)As nanowires have been grown either following a catalyst-free growth mode or using group-III catalyst droplets.
Catalyst-free vapour-solid growth of ordered (In,Ga)As NW arrays resulted in intermixed wurtzite (WZ)/zincblende (ZB) segments and thus a high density of stacking defects \cite{treuWidelyTunableAlloy2016}.
Self-assisted vapour-liquid-solid (VLS) growth using (In,Ga) droplets suffers from difficulties in achieving high In content in the solid ($x > 0.05$) \cite{heissCatalystfreeNanowiresAxial2009}. Only through radial shell growth, the In content could be increased to $x = 0.2$ near the surface \cite{heissGaQuantumDot2011}. Using an In-rich catalyst droplet, Scaccabarozzi \emph{et al.}~\cite{scaccabarozziStableHighYield2020} have incorporated an In fraction of $x \approx 0.2$ in the bulk of the NW, while improving the vertical yield of ordered NW arrays by $30 \%$---the main aim of the study---compared to GaAs NWs grown with a pure Ga catalyst droplet. The difficulties in achieving high $x$ in the NW bulk using the VLS growth mode have been theoretically substantiated by Dubrovskii~\cite{dubrovskiiUnderstandingVaporLiquid2017} and Scaccabarozzi \emph{et al.}~\cite{scaccabarozziStableHighYield2020}. 
	

	In this work, we report a detailed compositional, morphological and optical study of (In,Ga)As and (In,Ga)As/(In,Al)As core-shell NWs grown using binary (In,Ga) alloy droplets. The aim is to evaluate the limits to the In incorporation in the VLS growth mode, while achieving a high vertical yield without an extrinsic metal catalyst. NWs are grown by selective-area molecular beam epitaxy on Si(111) substrates. In particular, we investigate the effect of the elemental fluxes on the growth process, the morphology and the optical properties of the NWs.

	\section{Experiment}\label{experiment}
	
	\subsection{NW growth}\label{growth}

	All the In$_{x}$Ga$_{1-x}$As NWs of this study were grown on Si(111) substrates by MBE using the vapor-liquid-solid growth mode employing (In,Ga) alloy droplets. The substrates were covered with a thermal SiO$_{2}$ mask layer that was patterned by  electron-beam lithography to enhance the control and reproducibility of NW growth. Details can be found elsewhere \cite{kupersSurfacePreparationPatterning2017}. Initially, In and Ga were co-deposited at 590~\textdegree C for 30~s, using the same fluxes as for the subsequent growth. In this way, (In,Ga) alloy droplets were formed in the apertures of the SiO$_{2}$ mask \cite{scaccabarozziStableHighYield2020}. The (In,Ga)As growth was carried out for 30 min, while providing the samples with steady In, Ga and As fluxes at a substrate temperature of 590~\textdegree C. 
	
	To investigate the effects of the different elemental fluxes on the growth and properties of the (In,Ga)As NWs, we grew three separate series of samples. All the samples under study were grown with a constant sum of the group-III fluxes. (In,Ga)As NWs grown with different As fluxes (0.35, 0.44 and 0.52~ML/s) and a fixed In/(In+Ga) flux ratio of 0.50 form Series I. Series II consists of NWs grown with an As flux of 0.35 ML/s while changing the In/(In+Ga) flux ratio from 0.33 to 0.80. Lastly, Series III comprises additional NWs similar to Series II grown with In/(In+Ga) flux ratios of 0.5, 0.67 and 0.80 for an extended growth time of 60 min, where subsequently the droplet was consumed and a passivating (In,Al)As shell was grown. To consume the droplet, the group-III shutters were closed and the As flux was maintained for another 10 min. Thus, further (In,Ga)As formed, leading to a rounded shape of the NW top, as reported previously \cite{scaccabarozziStableHighYield2020}. The NW cores from this series were subsequently enclosed by a 10~nm thick (In,Al)As shell grown at 490~\textdegree C for 14~min, in order to passivate the surfaces and enhance the internal quantum efficiency in luminescence experiments \cite{jabeenRoomTemperatureLuminescent2008}. The flux ratio between In and Al during shell growth was adjusted according to measurements of the composition in equivalent NWs of Series II (see Results section). Hence, the lattice mismatch between shell and core is always minimal, whereas the bandgap is wider in the shell than in the core.

	\subsection{Characterization}\label{characterization}
	Energy dispersive X-ray (EDX) spectroscopy was used to quantitatively obtain the spatially-resolved solid In fraction $x$ of (In,Ga)As NWs. Measurements were acquired by either an EDAX Apollo XV detector or an EDAX Octane Elect EDS System mounted on a Zeiss Ultra55 field-emission scanning electron microscope (SEM) operated at 6.5~kV. The EDX quantification process was performed using the standardless eZAF algorithm provided by EDAX \cite{eggertEffectSiliconDrift2020}. The composition analysis of (In,Ga)As nanowires was complemented by the utilization of scanning transmission electron microscopy (STEM) with high-resolution energy-dispersive X-ray spectroscopy (EDX). Employing an aberration-corrected JEOL ARM 200F microscope and a Jeol Centurio 100 silicon drift detector, high angle annular dark-field (HAADF) micrographs and EDX spectra were simultaneously acquired. The microscope was operated at 200~keV using a collection angle of 54-220~mrad.
	
	For the purpose of studying the morphology of NWs from series II, further secondary electron (SE) micrographs were recorded with a Hitachi S-4800 at 5 kV and a tilt of 60\textdegree. These micrographs were analysed by an automated machine-learning algorithm based on the \textit{k-means clustering} method \cite{liClusteringMethodBased2012,steinleyKmeansClusteringHalfcentury2006}. NW arrays with a range of nominal hole openings ($d = 30$, 40, 50, 70 and 90 nm) and a fixed pitch of 1~\textmu m were studied. The length, diameter and vertical yield were obtained by averaging over many NWs for each of the NW fields. For example, 300 NWs were studied for fields with In/(In+Ga) of 0.33 while 50 NWs were studied for In/(In+Ga) of 0.80~\footnote{The developed code is available under \href{https://github.com/ACastMtz/SEM_ImgAnlysr.git}{this link}.}.
	

	
	
	Low-temperature cathodoluminescence (CL) and photoluminescence (PL) spectroscopy were used to explore the optical properties of the samples under study. CL line scans and monochromatic maps were acquired with a Gatan MonoCL4 system using a liquid nitrogen-cooled Si charge-coupled device (CCD) and a Peltier-cooled photomultiplier tube (PMT) as detectors, respectively. A spectral resolution of approximately 0.6~nm is achieved using a 300~mm focal length Czerny-Turner spectrometer with a 600 lines mm$^{-1}$ grating and slits of 0.1~mm. All the 10~K CL measurements were performed with a liquid-He-cooled stage using an SEM acceleration voltage of 5~kV and a beam-current of 0.94~nA. PL measurements were carried out at 10~K in a home-built PL setup with a continuous-wave (cw) Melles Griot 25 LHR 925 HeNe laser (632.8~nm) as excitation source with a spot diameter of 5~\textmu m and using a liquid-nitrogen cooled (In,Ga)As detector array. CL and PL experiments were performed on dispersed and as-grown NWs, respectively.
	
	\section{Results}
	
	\subsection{Composition and morphology}\label{influence_fluxes}
	
\begin{figure*}
		\centering
		\includegraphics*[width=\textwidth]{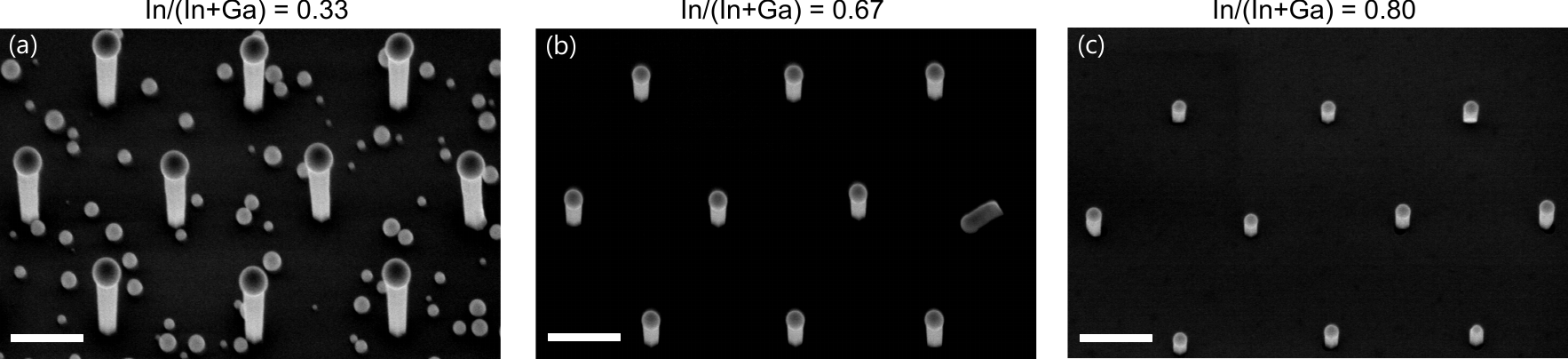}
		\caption{\label{sem01} Bird's eye view (60\textdegree) SE micrographs of (In,Ga)As NW arrays from Series II, grown with varying In/(In+Ga) flux ratios. The nominal hole sizes and pitch of the SiO$_{2}$ mask are 50 nm and 1 \textmu m, respectively. All the SE micrographs were acquired with the same magnification and the scale-bars are 500 nm long.}
	\end{figure*}

	We examined the impact of the In, Ga and As molecular beam fluxes on the compositional and morphological properties of (In,Ga)As NWs from Series I and~II. SE micrographs of (In,Ga)As NW arrays from Series II synthesized with different In/(In+Ga) flux ratios are presented in figure~\ref{sem01}. We observe the following behaviour: (i) NWs grow on the apertures of the pre-patterned SiO$_{2}$ layer, standing vertically on the surface of the Si(111) substrate. (ii) The catalyst-droplet remains intact on the NW tip once the (In,Ga)As growth has concluded. (iii) Both the length and diameter of the NWs decrease drastically when increasing the metal flux ratio, substantially reducing the NW volume.
	
\begin{figure}
		\centering
		\includegraphics[width=0.5\textwidth]{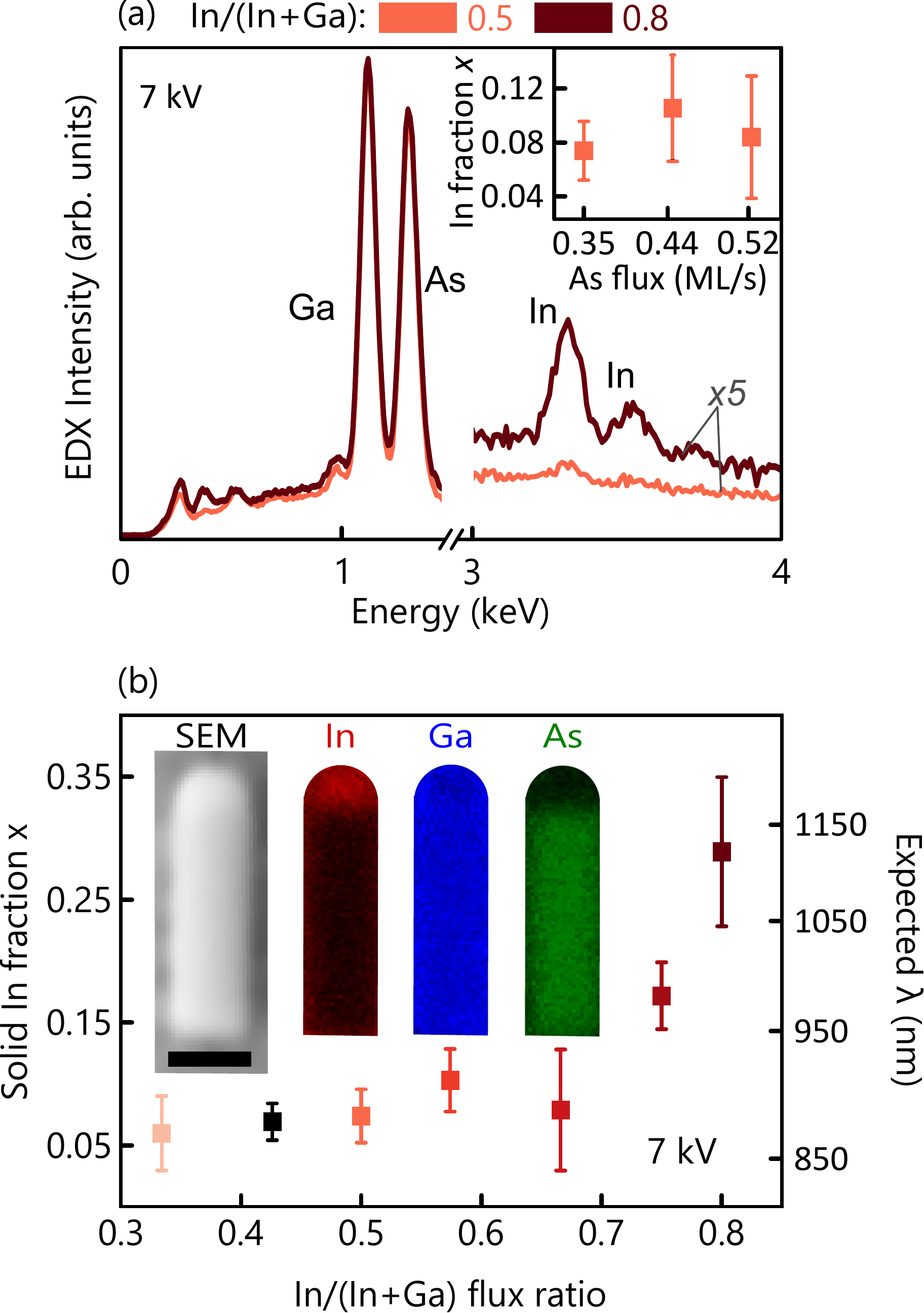}
		\caption{\label{edx01}  (a) Representative SEM-EDX normalised spectra of series II with In/(In+Ga) flux ratios of 0.5 and 0.8. The Ga and As peaks are formed by the superposition of the corresponding $L_{\alpha}$ and $L_{\beta 1}$ lines, while the In~$L_{\alpha}$ and In~$L_{\beta 1}$ lines are spectrally distinct. The inset shows the solid In fraction $x$ for Series I, with In/(In+Ga) of 0.5 and As fluxes of 0.35, 0.45, and 0.52 ML/s.~(b) Measured $x$ and corresponding emission wavelength $\lambda$ (right scale) for Series II, with varying In/(In+Ga) flux ratio. Quantitative SEM-EDX data in (a) and (b) were acquired with the electron beam impinging on the centre of individual NWs and averaged over multiple NWs. The inset to (b) shows SEM-EDX elemental maps and the corresponding SE micrograph for a single NW with As flux of 0.35~ML/s and In/(In+Ga) of 0.75. The scale-bar is 100 nm long. 
		}
	\end{figure}

	EDX was used to quantitatively obtain the solid In fraction $x$ of individual (In,Ga)As NWs. Figure~\ref{edx01}(a) shows characteristic spectra from Series II with In/(In+Ga) flux ratios of 0.50 and 0.80. The In/Ga peak intensity ratio is significantly increased for an In/(In+Ga) flux ratio of 0.80, directly revealing a higher In incorporation. The inset displays the measured In content for all three samples of Series I as a function of the As flux. It is clear that the As flux does not significantly affect $x$.
	
	In figure~\ref{edx01}(b), we present the measured $x$, and the corresponding expected band-to-band emission wavelength \cite{bhattacharyaPropertiesLatticematchedStrained1993a} $\lambda$ of NWs from Series II for different In/(In+Ga) flux ratios at an As flux of 0.35 ML/s. The In content $x$ in the solid equals $0.06 \pm 0.03$ for In/(In+Ga) of 0.33. There is no significant change in the NW composition for flux ratios up to 0.67. However, a notable increase of $x$ is obtained for higher flux ratios reaching $x=0.29\pm0.06$ at In/(In+Ga) of 0.80. The inset of figure~\ref{edx01}(b) displays the SEM image along with the corresponding EDX elemental maps of In, Ga and As for a single NW from the sample grown with a metal flux ratio of 0.75. From these maps, we can infer that the In content is significantly higher in the droplet at the NW tip than in the ternary NW. The droplet is also composed to a smaller extent of Ga, which in contrast is dominating the group-III content in the main NW segment. In particular, (In,Ga)As NWs grown with In/(In+Ga) of 0.5 exhibit an In content of $x=0.55\pm0.07$ in the catalyst droplet and $x=0.07\pm0.02$ in the solid NW. In conclusion, the system requires a high In flux in order to achieve a significant incorporation of this element into the NWs. This experimental finding agrees with previous measurements and theoretical expectations \cite{scaccabarozziStableHighYield2020}: Modelling the growth based on nucleation theory or standard equilibrium thermodynamics, it was found that the composition of the solid depends on that of the liquid in a highly nonlinear, threshold-like fashion. The In incorporation in the solid is predicted to be essentially zero up to an In fraction in the droplet of about 98\%. Above this threshold, the In incorporation should increase drastically over a small compositional range of the droplet. Finally, as expected, As is present only in the ternary NW and does not constitute a significant part of the droplet.

	We examined the impact of the group-III molecular beam fluxes on the NW morphology of series II by using an automated machine-learning image analysis. From this automated analysis, we obtain the values of the vertical yield ($\eta$), NW length and diameter as a function of the In/(In+Ga) flux ratio and the nominal hole sizes in the SiO$_{2}$ mask. Regardless of the growth parameters, with the exception of the sample grown with an In/(In+Ga) flux ratio of 0.50, the $\eta$ values fall between 60\% and 80\%. The reduced vertical yield of $\eta \approx 30$\% observed for In/(In+Ga) of 0.50 comes along with an increase of parasitic growth on the substrate. We attribute both effects to problems during the surface preparation process. A previous work reported an $\eta$ = (50 $\pm$ 20)\% for GaAs NWs grown with a Ga catalyst droplet, which increased to  $\eta$ = (80 $\pm$ 10)\% for (In,Ga)As NWs grown with an In catalyst droplet at an In/(In+Ga) flux ratio of 0.5 \cite{scaccabarozziStableHighYield2020}. 
	
\begin{figure}

		\includegraphics[width=\columnwidth]{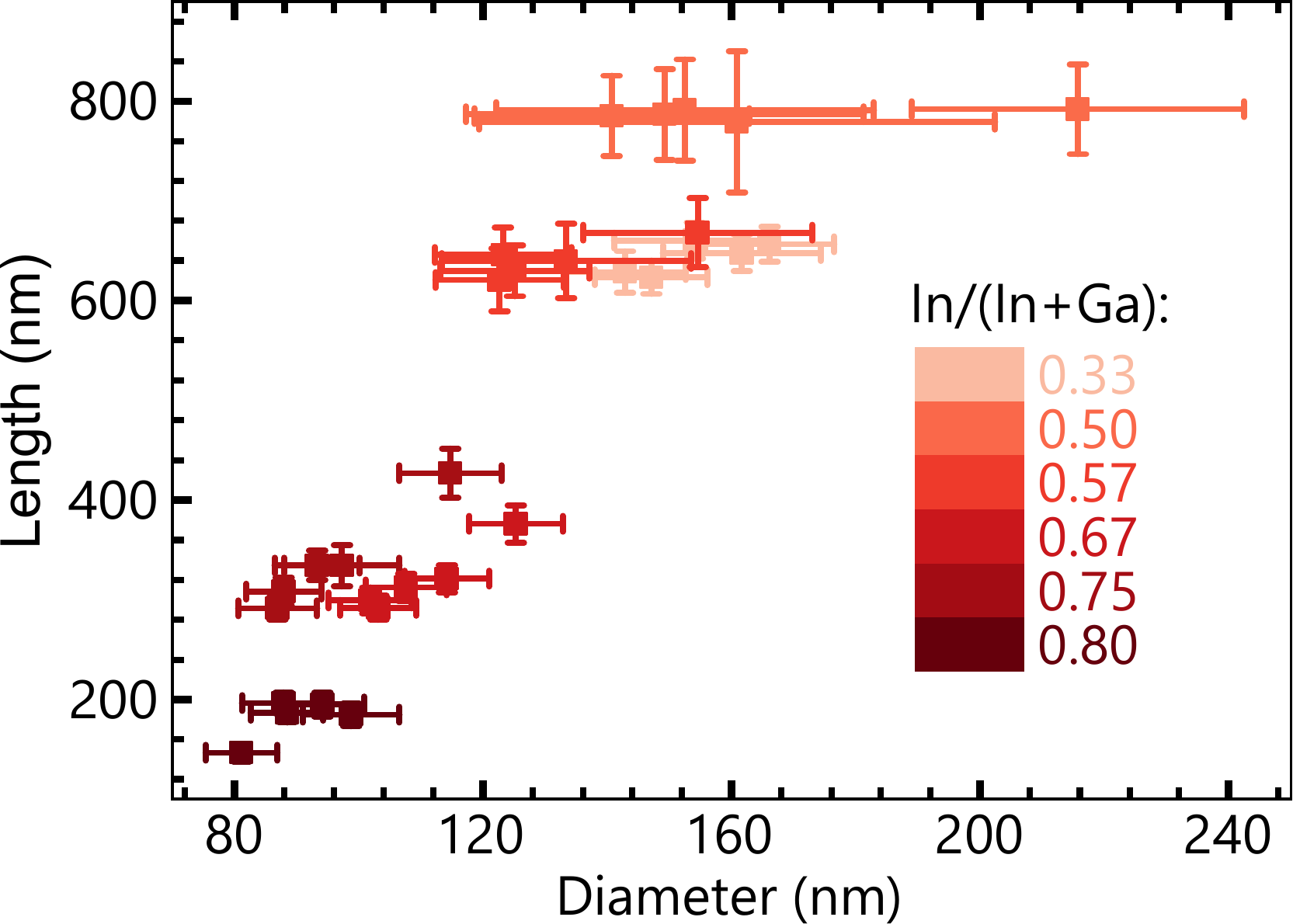}
		\centering
		\caption{Average NW length and diameter for different In/(In+Ga) flux ratios of series II, following the color code of Figure~\ref{edx01}(c). Different data points for each sample correspond to different nominal hole sizes of the SiO$_{2}$ mask: 30, 40, 50, 70 and 90 nm.
		}
		\label{fig:morphology}
	\end{figure}

	Figure ~\ref{fig:morphology} shows the measured length and diameter for different In/(In+Ga) flux ratios. The multiple data points per sample refer to different nominal aperture diameters in the SiO$_{2}$ film: $d = 30$, 40, 50, 70 and 90~nm. We observe that the NW diameter decreases from approximately $(160 \pm 30)$~nm to $(90 \pm 10)$~nm and that the length is equally reduced from $(640 \pm 20)$~nm to $(180 \pm 20)$~nm as the In/(In+Ga) flux ratio increases from 0.33 to 0.80. This drastic reduction of the NW volume by a factor of twelve is directly related to the Ga available for the growth. For samples with high In/(In+Ga) flux ratio, the Ga flux is lower, as the sum of the group-III material supplied by the molecular beams is always kept constant. The significant portion of In that is not incorporated into the NWs is presumably desorbed due to the high growth temperature utilized \cite{reithmaierIndiumDesorptionMBE1991}. The values for In/(In+Ga) of 0.50 are larger than expected, together with a larger dispersion of the values. We attribute this observation to the low vertical yield ($\eta\approx 30\%$) that degrades the performance of the automated analysis.

	As seen in figure~\ref{edx01}(b), a more effective incorporation of In is achieved for In/(In+Ga) flux ratios above 0.67. However, the data in figure~\ref{fig:morphology} indicate that the low liquid Ga fraction limits the growth process, reducing the NW dimensions as the supply of this element decreases. As a particular case, NWs with In/(In+Ga) ratio of 0.67 have their dimensions considerably reduced already, although there is no increase in the measured In content [figure~\ref{edx01}(b)]. The amount of Ga in the catalyst droplet is reduced to a degree that obstructs the growth and reduces the NW volume, but at the same time the In content is not yet high enough to incorporate In effectively.
	
	When comparing different hole diameters for a certain flux ratio, there is a slight reduction of the volume for arrays grown from smaller holes. NWs. However, this influence of the pre-patterned mask is much weaker than the one induced by the change in the In/(In+Ga) flux ratio. In fact, the variation of about 30~nm among the measured NW diameters for different nominal hole apertures at one flux ratio is in most of the samples lower than the range of nominal hole diameters of 30--90 nm.
	
	
	\subsection{Luminescence}\label{attribution}
\begin{figure}
		
		\includegraphics[width=0.50\textwidth]{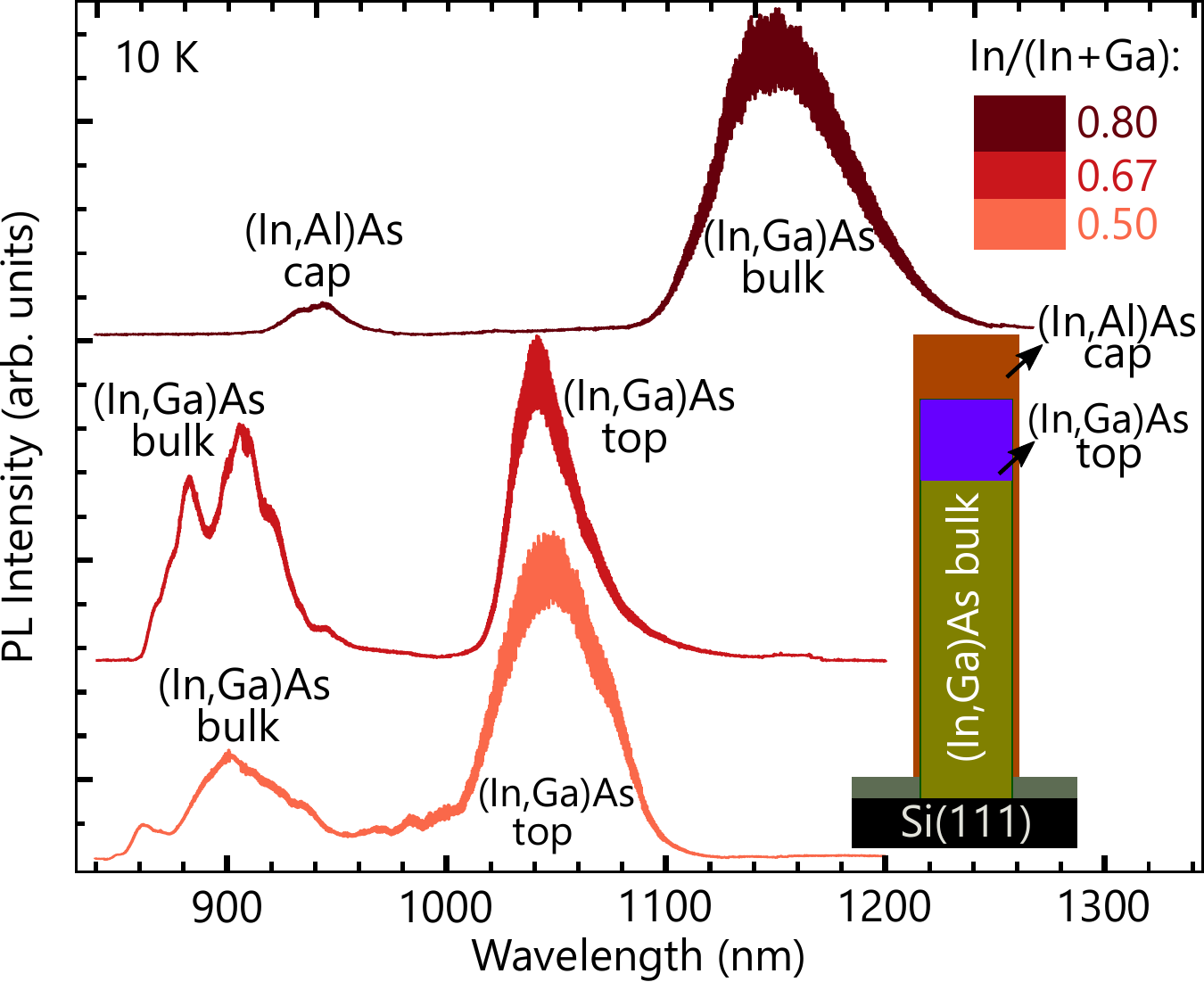}
		\caption{\label{PL_diffsamples} PL spectra at 10 K of (In,Ga)As/(In,Al)As core-shell NW ensembles from Series III with different In/(In+Ga) ratios. Spectra have been normalised and color-coded following Figure ~\ref{edx01}(b). The PL spectra shown in this figure were acquired with an excitation density of $10^{-2}I_{0}$. The inset shows a schematic of the regions of the NW that show luminescence represented in the spectra.}
	\end{figure}

	Figure \ref{PL_diffsamples} shows the PL spectra at 10~K for NW ensembles of series III with In/(In+Ga) ratios of 0.50, 0.67 and 0.80. Each of the spectra is composed of two emission bands, which points to differences in the In content, either between different NWs or within individual NWs, as the bandgap of strain free (In,Ga)As is directly related to $x$ \cite{bhattacharyaPropertiesLatticematchedStrained1993a}. The different parts of the NWs that give rise to the spectrally separated luminescence bands are shown as elucidated below in the inset: (In,Ga)As bulk, (In,Ga)As top segment generated by the consumption of the metal droplet and the axially grown (In,Al)As cap deposited during shell growth. We also performed corresponding room-temperature (300~K) PL experiments in these same samples (data not shown). The non-quenching of the PL emission at 300~K indicates the great potential of these (In,Ga)As NWs for optoelectronic applications.
	
	The emission bands do not shift between the NW ensembles with In/(In+Ga) of 0.50 and 0.67, in agreement with the observation of a similar In content in the equivalent samples of Series II. In particular, the emission at 900~nm matches well with $x=0.08 \pm 0.05$ measured by SEM-EDX for the corresponding samples from Series II (figure \ref{edx01}(b)). The spectrum for the sample with an In/(In+Ga) flux ratio of 0.80 shows two PL bands at longer wavelengths. Following the same reasoning as before, the dominating emission at 1150~nm for In/(In+Ga) of 0.80 agrees with $x=0.29 \pm 0.06$ obtained from SEM-EDX. The origin of the second emission band observed for each sample is less obvious. In the particular case of the sample grown with an In/(In+Ga) flux ratio of 0.8, the spectrum does not show a red-shifted emission as for the other samples, but rather a blue-shifted secondary emission.
	
		\begin{figure*}
		
		\includegraphics*[width=1\textwidth]{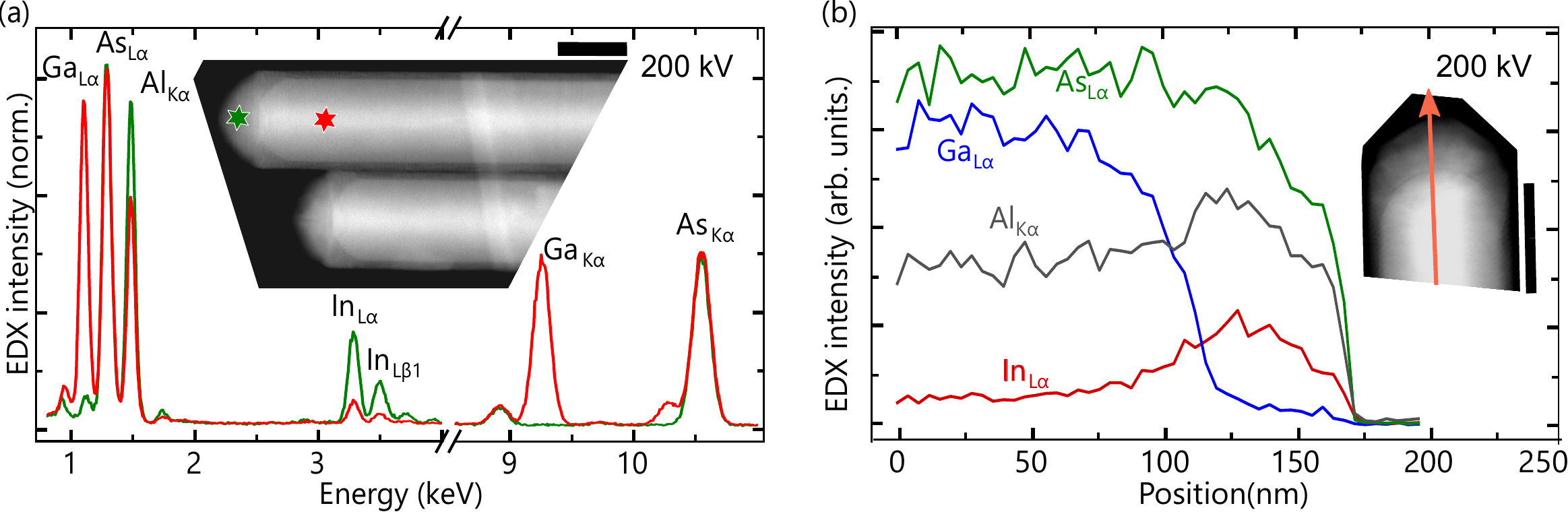}
		\caption{\label{EDX_tem} (a) Normalised STEM-EDX spectra of a (In,Ga)As/(In,Al)As core-shell NW of series III grown with In/(In+Ga) = 0.8 dispersed on a TEM grid. The inset shows the corresponding HAADF-STEM image and the position of the electron beam is indicated. (b) STEM-EDX elemental distribution of In, Ga, As and Al of a (In,Ga)As/(In,Al)As core-shell NW of series III grown with In/(In+Ga) = 0.8 along the red line shown in the inset HAADF-STEM image. Both scale bars shown in this figure are 100 nm. }
	\end{figure*}

		In order to clarify the spatial origin of the emission band centred at 940~nm observed for NWs grown with In/(In+Ga)~=~0.8 (figure~\ref{PL_diffsamples}), we performed STEM-EDX measurements. Figure~\ref{EDX_tem}(a) shows representative STEM-EDX spectra recorded at the centre of the (In,Ga)As core and the (In,Al)As cap as indicated in the corresponding micrograph. We distinguish the (In,Ga)As core, (In,Ga)As top generated by the consumption of the metal droplet, (In,Al)As shell and the (In,Al)As cap in this micrograph. From this HAADF-STEM image we estimate the (In,Al)As cap to be approximately 50 nm long whereas the shell thickness is equal to the nominal 10 nm. From the spectra shown in Figure~\ref{EDX_tem}(a), the In signal intensity from the (In,Al)As cap is more than doubled compared with the (In,Ga)As core. The Al signal in the core spectrum originates from the lateral shell. Figure~\ref{EDX_tem}(b) displays the EDX intensity profile for various elements captured along the red line indicated in the inset, confirming a significant increase in the In signal within the (In,Al)As cap, coupled with a reduction in the Ga signal. The quantitative SEM-EDX on NW cores in Figure 2 resulted in an In content of $0.29\pm0.06$ for samples grown at this flux ratio, whereby we estimate an In content of at least 0.6 in the (Al,In)As cap. The corresponding emission wavelength for a uniform alloy would be 940~nm. Therefore, we attribute the weak PL band at this wavelength observed in figure \ref{PL_diffsamples} to the axially grown (In,Al)As cap.

\begin{figure*}
		\includegraphics*[width=\textwidth]{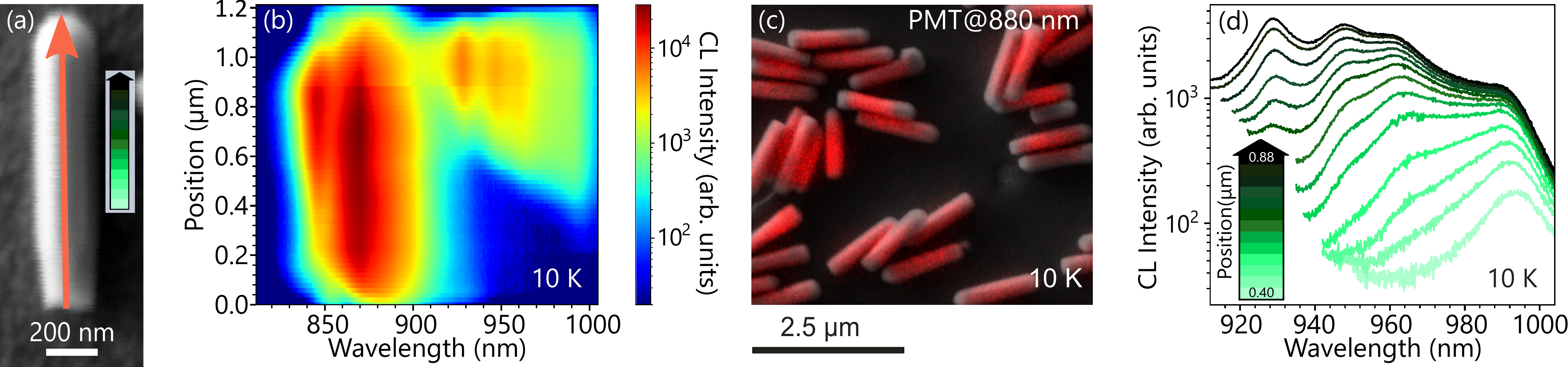}
		\caption{\label{CL} (a) SE micrograph of a single (In,Ga)As/(In,Al)As core-shell NW of Series III with an In/(In+Ga) flux ratio of 0.50. (b) Corresponding CL spectral line scan along the NW axis, as indicated by the arrow. CL intensity is plotted on a logarithmic, color-coded scale. (c) SE micrograph overlaid by a monochromatic CL image recorded at 880 nm. (d) CL spectra for the long wavelength band extracted from the line scan in (a) at different positions indicated by the green arrow in (a). All the data shown in this figure were acquired at 10 K.}
	\end{figure*}

	To further investigate the spatial origins of the different luminescence bands, we performed CL experiments on single NWs, as exemplified in figure ~\ref{CL}. Figures~\ref{CL}(a) and (b) show an SE micrograph of a NW grown with an In/(In+Ga) flux ratio of 0.5 and the corresponding CL spectral line scan  performed along the axis of the NW (as indicated by the red arrow) recorded at 10~K, respectively. The line scan reveals two different emission bands, as expected from the PL measurements (figure~\ref{PL_diffsamples}). The first band, centered around 880~nm, is present along the entire NW, but weakened towards the tip. This emission clearly corresponds to the short wavelength band observed from PL for this sample. As confirmation, figure \ref{CL}(c) presents an SE micrograph of multiple NWs superimposed by the monochromatic PMT image obtained at 880~nm. We infer that this emission is present for all NWs over most of their length, but usually is missing towards the tip of the NWs (rounded end). 

	Near the NW tip, where the growth concluded, we observe a second emission at longer wavelengths that we associate with the long wavelength band seen in the PL spectra (figure~\ref{PL_diffsamples}). The discrepancy in emission wavelength is clarified below. From the SEM-EDX maps shown in figure~\ref{edx01}(b), In is the predominant element in the droplet after the (In,Ga)As growth concluded. Consequently, when only keeping the As flux in order to consume the latter, the In incorporation is enhanced. The In-rich NW tip causes the redshifted emission peaks. The emission from the (In,Ga)As top for In/(In+Ga)~=~0.8 may also be present, but the emission wavelength is expected to lie outside of our detection range. The origin of the multiple peaks observed in this region of the NW will be discussed below.

	
	From the CL line scan in figure~\ref{CL}, we further notice that the emission from the (In,Ga)As top extends over 600~nm in length, whereas we expect the corresponding segment together with the (In,Al)As cap to be approximately 200-300~nm long. We estimate the axially grown (In,Al)As cap to be 50~nm long, as observed in the HAADF micrographs (figure~\ref{EDX_tem}) and in the upper dark segment from from figure~\ref{CL}(b). In figure~\ref{CL}(d), we show CL spectra of the long wavelength emission extracted from figure~\ref{CL}(b) at the positions indicated by the green arrow in figure~\ref{CL}(a). At the spatial onset of this emission band (vertical position of 0.4~\textmu m), when the electron beam is far from the top segment and the excitation density is low, the band is formed by a single peak centred at 990~nm. As the electron beam is scanned towards the NW tip and the excitation density in that region increases, further peaks appear at shorter wavelengths (higher energies): 963, 947 and 929~nm. We attribute both phenomena to the inherent limits of the spatial resolution in CL and the peculiarity that the excitation is local, but the detection is from a larger area \cite{edwardsCathodoluminescenceNanocharacterizationSemiconductors2011}. The long carrier diffusion lengths reported for (In,Ga)As \cite{niemeyerMinorityCarrierDiffusion2017a,gustafssonDeterminationDiffusionLengths2010a} in combination with the carrier generation volume relevant for CL explain why emission from the top segment is already observed when the beam is far from this segment. Under the employed SEM acceleration voltage (5 kV) and measurement temperature (10 K), the generation volume relevant for CL has a radius of approximately 100~nm \cite{jahnCarrierDiffusionGaN2022a}. At the same time, when the top segment is only indirectly excited by diffusion, the excitation density in this segment is lower. The origin of the multiple peaks observed in figure~\ref{CL}(d) is discussed below.

\begin{figure}
		\centering
		\includegraphics*[width=0.4\textwidth]{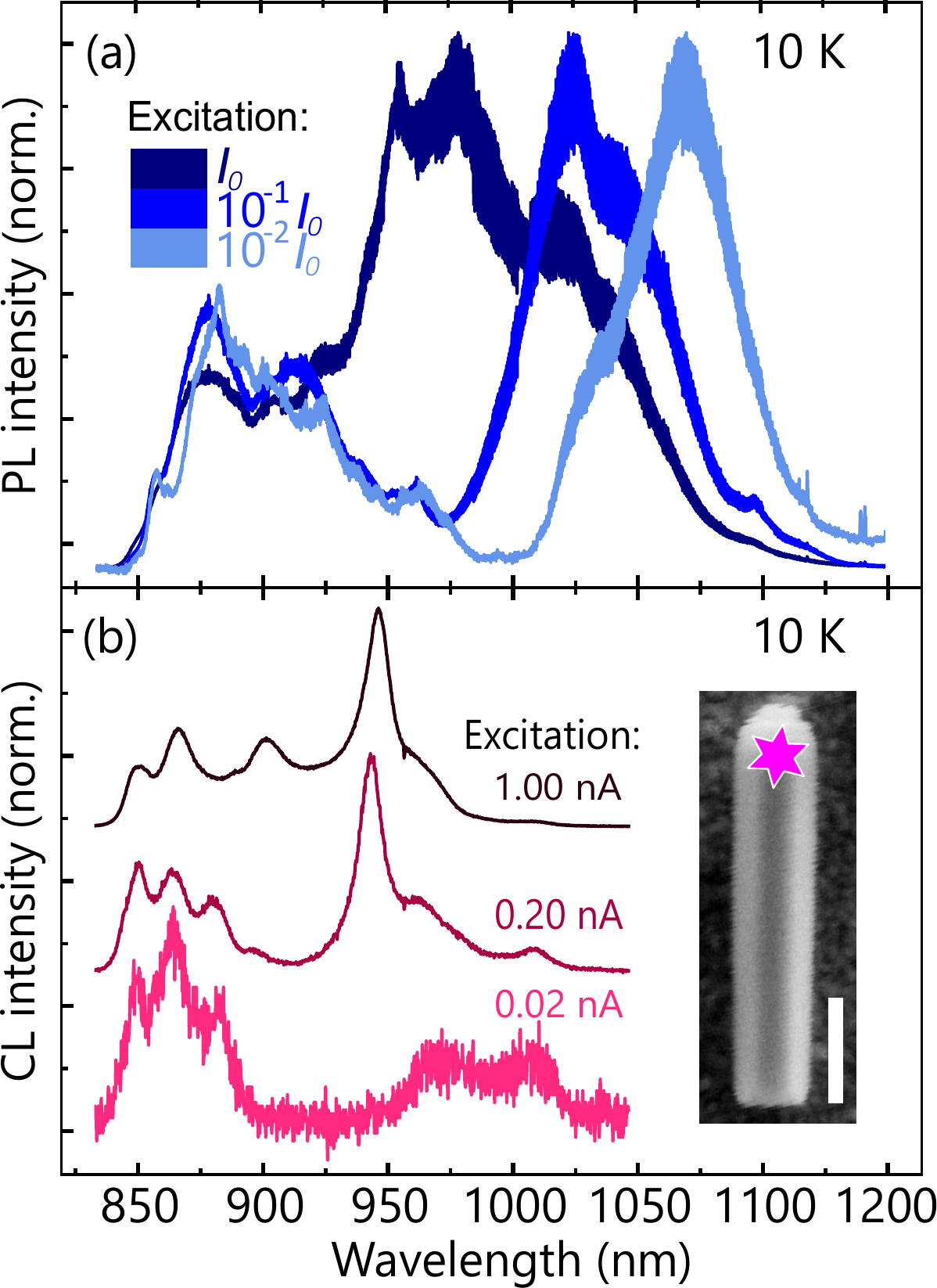}
		\caption{\label{excitation}(a) Normalised cw-PL spectra at 10~K of (In,Ga)As/(In,Al)As core-shell NW ensembles from series III with In/(In+Ga) = 0.5 and different excitation densities. (b) CL spectra recorded with different beam currents with the electron beam focused on the tip of a NW from the same ensemble. The SE micrograph of the inset has a 500~nm scalebar. The position of the electron beam is indicated.}
	\end{figure}

	To further study the (In,Ga)As top band, we turn to excitation-dependent PL experiments. Figure~\ref{excitation}(a) depicts PL spectra of (In,Ga)As/(In,Al)As core-shell NW ensembles with an In/(In+Ga) flux ratio of 0.5 obtained at 10~K for excitation densities varied over two orders of magnitude. The bulk emission at around 880~nm is not shifted with increasing excitation. In the same excitation range, the top emission shows a notable blue-shift of around 100~nm. This excitation dependence of the top emission explains the discrepancy between PL and CL in figure~\ref{PL_diffsamples} and figure~\ref{CL}, where a higher excitation density was used in the CL measurements. Equivalent PL experiments performed in InAs \cite{koblmullerDiameterDependentOptical2012} and In$_{0.81}$Ga$_{0.19}$As \cite{morkotterRoleMicrostructureOptical2013} NW ensembles show a much weaker blue-shift along similar excitation density ranges that was attributed to the filling of conduction band states with degenerate non-equilibrium electrons \cite{arnaudovBandfillingEffectLight2009}. To clarify the actual origin of the blue-shift and the appearance of multiple luminescence peaks on the (In,Ga)As top, we further varied the SEM beam currents during CL the measurements.
	
	Figure~\ref{excitation}(b) shows CL spectra recorded at 10~K with different beam currents and thus excitation densities on a NW from series III with an In/(In+Ga) flux ratio of 0.5. The electron beam was impinging on the tip of the NW, as indicated in the inset. For a beam current of 0.02~nA, the top emission consists of two different, weak peaks centred at 970~nm and 1009~nm. As the electron beam current is increased towards 0.20~nA and 1~nA, the latter peaks are still present while additional, more intense peaks emerge at 943 and 910~nm. This excitation dependence corresponds to the same effect seen in figure~\ref{CL}(c), where subsequent CL peaks of the top emission appear when the electron beam gets closer to the NW tip. 
	
	We understand the complex CL spectra of the (In,Ga)As NW tip as a consequence of a series of localized states with limited density of states, where only the lowest energy states are contributing to the low-excitation spectra. With saturation of these lowest energy states at increasing excitation, higher energy peaks appear. Such localized states may be generated when the droplet is consumed under a high group-V flux and could be related to compositional disorder of the ternary alloy giving rise to clusters of higher In content. Another possible origin of the localized states could reside in alternating segments of ZB and WZ phases, i.e.\ a crystal structure with stacking defects, commonly reported for the droplet consumption region in GaAs NWs \cite{dastjerdiMethodsGaDroplet2016,lahnemannCorrelatedNanoscaleAnalysis2019}. Most likely, both effects contribute to the observed carrier localization. As a summary, the different regions of the NWs that show luminescence bands are the (In,Ga)As bulk, (In,Ga)As top segment generated by the consumption of the metal droplet and the axially grown (In,Al)As cap deposited during shell growth.

\begin{figure}
		\centering
		\includegraphics*[width=0.35\textwidth]{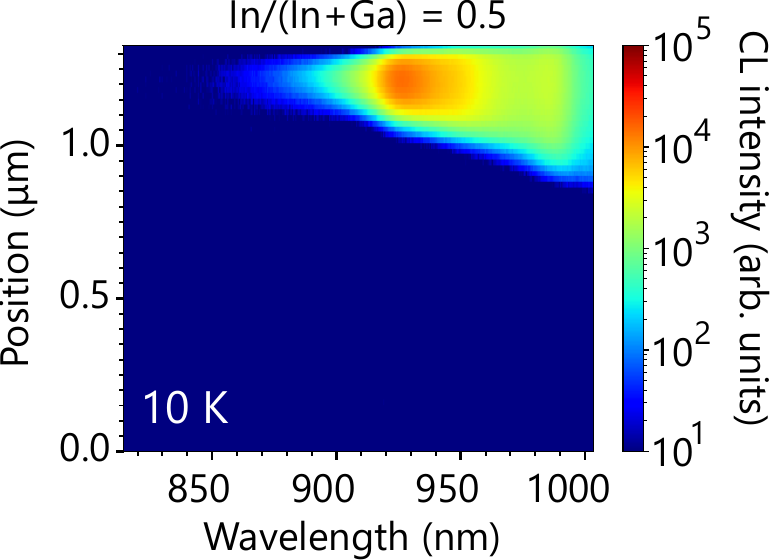}
		\caption{\label{oxidation}
			CL spectral line scan along the axis of a (In,Ga)As/(In,Al)As core-shell NW of series III grown with In/(In+Ga) = 0.5. Data was collected more than six months after the NWs were grown.}
	\end{figure}

	
	All the presented luminescence data so far was acquired within one month after the samples were grown. Additional CL experiments were performed more  than half a year after the (In,Ga)As/(In,Al)As core-shell NWs were grown. Figure~\ref{oxidation} shows a CL spectral line scan of a single NW of series III with an In/(In+Ga) flux ratio of 0.5 recorded at 10~K -- the same sample as shown in figure~\ref{CL}(a). We observed that all NWs exhibited a partial or total decrease of the bulk emission CL band, regardless of the growth conditions. We attribute this loss of emission to the oxidation of the (In,Al)As shell. The CL signal from the droplet consumption region is still observed, because localized states are decoupled from the surface. In particular, In-rich clusters located inside the NW and the type-II band gap alignment induced by the alternating ZB and WZ phases are independent of the surface oxidation. 
	
	Using GaAs instead of (In,Al)As as a shell around the (In,Ga)As core would have prevented the oxidation of the shell, but the strain induced by this shell would generate radial piezoelectric fields that would separate carriers \cite{moratisStrainedGaAsInGaAs2016} and therefore would most likely degrade the optical properties. To prevent these effects, a shell of (In,Al)As lattice-matched to the (In,Ga)As NW core would have to be combined with an additional $\approx$ 5~nm thin GaAs shell to prevent oxidation.


	\section{Conclusions}
	In summary, our study evaluates the limits of incorporating In in (In,Ga)As NWs grown by MBE using the self-assisted growth mode. Our results confirm the difficulty of incorporating a considerable amount of In, unless a high In/(In+Ga) flux ratio is employed. However, we were able to successfully elevate the In content to $0.29\pm0.06$ for an In/(In+Ga) flux ratio of 0.80. We report that the NW volume is reduced by a factor of twelve after increasing the In/(In+Ga) flux ratio from 0.33 to 0.80, while keeping the sum of metal fluxes constant. To achieve larger NWs, the total fluxes or the growth time would have to be increased accordingly. Towards the upper part, individual NWs show inhomogeneities in $x$ and crystal defects related to the consumption of the droplet that lead to complex band profiles strongly affecting their optical properties. Spatially-resolved as well as excitation-dependent measurements of the luminescence of (In,Ga)As/(In,Al)As core-shell NWs show a uniform emission over most of the NW length, where the emission wavelength agrees with the composition determined by SEM-EDX measurements on corresponding (In,Ga)As NWs without shell. This (In,Ga)As segment with uniform emission makes the studied NWs potential candidates for a site-selective integration of infrared emitters on Si platforms, in particular on waveguides.

	\ack
	The authors acknowledge XXX for a critical reading of the manuscript and Anne-Kathrin Bluhm for the SE micrographs. Furthermore, the authors are grateful for funding from the German Federal Ministry of Education and Research BMBF in the framework of project MILAS.

	\bibliography{InGaAs_MG_2023}
	
	
\end{document}